\documentclass[%
 aip,
 amsmath,amssymb,
 reprint
]{revtex4-1}

\usepackage{graphicx}
\usepackage{dcolumn}
\usepackage{bm}
\usepackage{xcolor}
\usepackage[utf8]{inputenc}
\usepackage[T1]{fontenc}
\usepackage{mathptmx}

\begin{document}

\preprint{AIP-AgBacteria}

\title[The Impact of Bacteria Exposure on the Plasmonic Response of Silver Nanostructured Surfaces]{The Impact of Bacteria Exposure on the Plasmonic Response of Silver Nanostructured Surfaces}

\author{Giuseppe M. Patern\`{o}}
\thanks{These authors contributed equally to this work.}
\affiliation{ 
Center for Nano Science and Technology, Istituto Italiano di Tecnologia (IIT), Via Pascoli 10, 20133, Milano, Italy
}
\author{Aaron M. Ross}%
\thanks{These authors contributed equally to this work.}
\affiliation{ 
Physics Department, Politecnico di Milano, Piazza L. da Vinci 32, 20133 Milano, Italy
}%
\author{Silvia M. Pietralunga}
\affiliation{ 
Center for Nano Science and Technology, Istituto Italiano di Tecnologia (IIT), Via Pascoli 10, 20133, Milano, Italy
}
\affiliation{ 
Institute for Photonics and Nanotechnologies (IFN), Consiglio Nazionale delle Ricerche (CNR), Piazza L. da Vinci 32, 20133 Milano, Italy
}
\author{Simone Normani}
\affiliation{ 
Center for Nano Science and Technology, Istituto Italiano di Tecnologia (IIT), Via Pascoli 10, 20133, Milano, Italy
}
\author{Nicholas Dalla Vedova}
\affiliation{ 
Center for Nano Science and Technology, Istituto Italiano di Tecnologia (IIT), Via Pascoli 10, 20133, Milano, Italy
}
\author{Jakkarin Limwongyut}
\affiliation{ 
Center for Polymers and Organic Solids, Department of Chemistry and Biochemistry, University of California, Santa Barbara, CA 93106, USA
}
\author{Gaia Bondelli}
\affiliation{ 
Center for Nano Science and Technology, Istituto Italiano di Tecnologia (IIT), Via Pascoli 10, 20133, Milano, Italy
}
\affiliation{ 
Physics Department, Politecnico di Milano, Piazza L. da Vinci 32, 20133 Milano, Italy
}%
\author{Liliana Moscardi}
\affiliation{ 
Center for Nano Science and Technology, Istituto Italiano di Tecnologia (IIT), Via Pascoli 10, 20133, Milano, Italy
}
\affiliation{ 
Physics Department, Politecnico di Milano, Piazza L. da Vinci 32, 20133 Milano, Italy
}%
\author{Guillermo C. Bazan}
\affiliation{ 
Center for Polymers and Organic Solids, Department of Chemistry and Biochemistry, University of California, Santa Barbara, CA 93106, USA
}
\affiliation{ 
School of Chemical and Biomedical Engineering Nanyang Technological University Singapore 639798, Singapore
}
\affiliation{ 
Singapore Centre on Environmental Life Sciences Engineering Nanyang Technological University Singapore 639798, Singapore
}
\author{Francesco Scotognella}
\email[Authors to whom correspondence should be addressed: ]{francesco.scotognella@polimi.it, guglielmo.lanzani@iit.it}
\affiliation{ 
Center for Nano Science and Technology, Istituto Italiano di Tecnologia (IIT), Via Pascoli 10, 20133, Milano, Italy
}
\affiliation{ 
Physics Department, Politecnico di Milano, Piazza L. da Vinci 32, 20133 Milano, Italy
}%
\author{Guglielmo Lanzani}
\email[Authors to whom correspondence should be addressed: ]{francesco.scotognella@polimi.it, guglielmo.lanzani@iit.it}
\affiliation{ 
Center for Nano Science and Technology, Istituto Italiano di Tecnologia (IIT), Via Pascoli 10, 20133, Milano, Italy
}
\affiliation{ 
Physics Department, Politecnico di Milano, Piazza L. da Vinci 32, 20133 Milano, Italy
}%

\date{\today}
\begin{abstract}
Silver, especially in the form of nanostructures, is widely employed as an antimicrobial agent in a large range of commercial products. The origin of the biocidal mechanism has been elucidated in the last decades, and most likely originates from silver cation release due to oxidative dissolution followed by cellular uptake of silver ions, a process that causes a severe disruption of bacterial metabolism and eventually leads to eradication. Despite the large number of works dealing with the effects of nanosilver shape/size on the antibacterial mechanism and on the (bio)physical chemistry pathways that drive bacterial eradication, little effort has been devoted to the investigation of the silver NPs plasmon response upon interaction with bacteria. 
Here we present a detailed investigation of the bacteria-induced changes of the plasmon spectral and dynamical features after exposure to one of the most studied bacterial models, \textit{Escherichia Coli}. Ultrafast pump-probe measurements indicate that the dramatic changes on particle size/shape and crystallinity, which stem from a bacteria-induced oxidative dissolution process, translate into a clear modification of the plasmon spectral and dynamical features. This study may open innovative new avenues in the field of biophysics of bio-responsive materials, with the aim of providing new and reliable biophysical signatures of the interaction of these materials with complex biological environments. 
\end{abstract}
\maketitle

\section{Introduction}

Silver-containing materials have been employed empirically for millennia as disinfectant agents \cite{Chernousova2013}. The mechanism of its antibacterial activity was elucidated in the 19th century, and is attributed to the toxicant effect of silver ions \cite{Xiu2012}, which are released via an oxidative dissolution process \cite{LeOuay2015}. Specifically, silver nanostructures are by and large the most commonly used form of materials for antibacterial purposes, mostly due to the high surface-to-volume ratio and, hence, to their relatively high bio-responsivity. Following the increase of antibiotic-resistant bacterial strains, the use of silver for disinfection, i.e. as antibacterial coatings, has seen a steep growth \cite{Chernousova2013}. Accordingly, the number of research and review articles on the biocidal effect of nanosilver materials has seen a substantial rise, with most of the works dealing with the biological effects leading to bacteria eradication \cite{Chen2000,Lemire2013,Rizzello2014} or with the physical chemistry aspects that govern its biocidal action \cite{LeOuay2015}. 

However, the changes in the optical properties of silver (Ag) upon interaction with bacteria are not well understood, despite the fact that silver nanostructures display well-established optical features arising from the localized surface plasmon resonance (LSPR) \cite{Cazalilla2000}. It is interesting to note that shifts in the Ag LSPR spectral position have been already used as a parameter to monitor the occurrence of oxidative dissolution of silver \cite{Mogensen2014}, which incidentally is a reaction pathway involved in the silver biocidal mechanism. In this regard, we have preliminarily investigated the modification of the Ag LSPR peak position after exposure to whole bacterial cells \cite{Paterno2019a, D0FD00026D, Normani2020}, with the aim to exploit such an effect for development of plasmonic/photonic colorimetric sensors of bacterial contaminants \cite{Paterno2020}. In particular, we have observed a blue shift of the plasmon response upon bacterial contamination in both linear absorption and pump-probe spectroscopy, a result that we have qualitatively attributed to an increase of the Ag free electron density upon bacteria/Ag interaction, as well as particle structural modification via oxidative dissolution.

Here, we present a systematic study on the impact of \textit{Escherichia Coli (E.coli)} exposure on the plasmon resonance of Ag nanoplate films. We found that exposure of the Ag films to \textit{E.coli} cells leads to particle amorphization and agglomeration in close proximity to the bacterial cells, as well as a marked change on the NPs shape. Pump-probe spectroscopy reveals that all those effects can act concurrently to yield a markedly different scenario in terms of plasmon frequency and relaxation dynamics, namely: i. a clear blue-shift of the LSPR peak that may stem from particles rounding-out and shrinking likely due to the oxidation process; ii. a marked decrease in the electron-phonon coupling time that we attribute to increased Ag free electron density;  iii. a damping of the coherent oscillations that can be related to the amorphization process. Interestingly, these changes in the plasmon spectra and dynamics are enhanced when \textit{E.coli} cells are treated with a plasma membrane permeabilizer before exposure to nanosilver, indicating that ionic release and uptake are the leading processes that ultimately drive the modification of nanosilver morphology and, hence, of its optical features. The paper is organized as follows: after a brief review of the possible processes affecting silver NPs chemical state and morphology upon interaction with bacteria (Section \ref{sec:lit}), we will be discussing the structural and morphological data, as well as the linear absorption spectra (Section \ref{sec:AgCharacterization}). The detailed investigation on the plasmon spectral and dynamical response upon \textit{E.coli} contamination carried out via ultrafast pump-probe is presented in Section \ref{sec:PP}. Finally, conclusions and future directions of our work are contained in Section \ref{sec:conclusions}.

\section{Literature review of interactions between Ag and \textit{E.coli}}\label{sec:lit}

Although the antimicrobial properties of silver have been well-known for millennia and still exploited in a wide range of everyday life applications, the actual mechanism underpinning such biocidal properties has only been deeply investigated in the last few decades. The first step of the process consists in the release of the positively charged silver ions (Ag+) and their transmembrane cell penetration \cite{Dakal2016}, which is most likely driven by the electrostatic interaction with the negatively charged bacterial cell wall \cite{ElBadawy2011}. The complex chain of events leading to bacterial eradication after ionic uptake might involve the production of reactive oxygen species (oxidative stress), protein dysfunction and membrane damage, and are documented in some detailed reviews \cite{Stark2011, Lemire2013}.  

Despite the large wealth of works investigating any particle-specific effects on the antibacterial action of silver nanostructures, such as shape and size \cite{Rai2009, Pal2007}, Alvarez and collaborators have reported that Ag$^+$  is  the actual toxicant species, with other geometrical properties being indirect effectors that primarily influence Ag$^+$ release \cite{Xiu2012}. The origin of the oxidation of zerovalent silver and release of Ag$^+$ into solution has been attributed to the presence of molecular oxygen dissolved in the medium. Briefly, silver as a noble metal does not dissolve either in water or acids (E$_0$ = + 0.80 V), while the presence of an oxidizer such as molecular oxygen (E$_0$ = + 1.23 V) can lead to silver oxidation and release of Ag$^+$. Such a reaction can be enhanced by complexation of silver with nucleophilic agents that may possibly be present in the bacterial cells \cite{Chernousova2013}, such as organic thiolates. Note that nucleophiles are widely employed to enhance gold dissolution and extraction from metal ore via the use of cyanide as nucleophilic reagent (gold cyanidation). The reaction scheme for the oxidative dissolution would be:

\begin{equation}\label{eq:reaction}
Ag_m^0N + O_2 \to Ag_{m-1}^0 + Ag^+N + O_2^{\cdot -}
\end{equation}

where N is the nucleophilic ligand coordinated to the metal and O$_2^{\cdot -}$ is the superoxide radical \cite{Mogensen2014}. Notably, particle dissolution has been seen to impact greatly the morphology of silver NPs, with a rounding-out of the shape (i.e. from triangles to circles) and a decrease of the radius (about 5-10 nm) \cite{Kent2012}, in analogy with the rounding-out of particles observed during electrochemical oxidation of silver nanoplates \cite{Zhang2005}. However, the nucleophilic-mediated oxidative dissolution and the relative morphological effects on Ag NPs have been studied quantitatively only in controlled chemical/physical environments, while direct investigation over all the aforementioned effects has never been performed in the presence of actual bacterial cells. 

\section{Ag nanoplate (NP) thin film characterization, before and after \textit{E.coli} exposure}\label{sec:AgCharacterization}

\subsection{Experimental methods}\label{subsec:fabrication}

Silver films used in our studies consist of a set of nanoplates with an average sizes of 18 nm (standard deviation 6 nm, see Figure S1), obtained via thermal evaporation (rate = 0.01 nm/s, average thickness 8 nm) on glass substrates (MBRAUN metal evaporator).

For bacterial colony growth and exposure to silver surfaces, a single colony from the Escherichia coli ATCC25922 strain was inoculated in Luria-Bertani (LB) broth and incubated overnight at 37ºC with shaking at 200 rpm until stationary phase was reached. Then, bacterial suspension turbidity (expressed as optical density at 600nm; O.D.$_{600}$) was diluted to O.D.$_{600}$ ~ 0.5 (corresponding to 1.12x10$^7$ colony-forming unit (cfu)/ml) in LB broth (no antibiotic). The suspension (100 $\mu$L) was spread over an LB agar plate. The samples were then placed at the center of the Petri dish with the silver layer facing the contaminated surface (or LB only for the control experiment) and incubated for 24 h at 37 C°. To assess the role of silver ionic uptake in the modification of its plasmon response, we also treated the bacterial culture with a conjugated oligoelectrolyte (COE2-3C, see Figure S2 for the chemical structure) that is known to enhance the permeability of bacterial membrane. The properties of this class of molecules, alongside their synthesis can be found in a recent paper \cite{Limwongyut2020}. For this experiment, COE2-3C was mixed with the bacterial suspension to obtain a final concentration of 20 $\mu$M. This \textit{E.coli}/COE mix in LB broth was then incubated overnight  at 37 C° with shaking at 200 rpm, following the same procedure used for \textit{E.coli} cells culture. 

SEM images were taken using a Tescan MIRA3 High-RES SEM. Secondary Electron imaging was performed at a voltage of 5 kV and with currents in the pA range. Silver layers and bacteria were deposited on top of silicon p-doped substrates and electrically contacted with carbon tape to avoid electrostatic charging.

X-ray diffraction measurements were carried out with a Bruker D8 advance in a grazing-angle thin film 2$\theta$ scan model geometry ($\alpha$ = 0.5$^{\circ}$), by using a K$\alpha$ wavelength emitted by a Cu anode (0.15418 nm, 40 kV). 

The optical absorption was recorded using a fiber-coupled spectrometer (Avantes, AvaSpec-HS2048XL-EVO) equipped with a deuterium-halogen lamp (AvaLight-D(H)-S). Absorption spectra were corrected for blank (bare glass substrate) and dark signal. We took a minimum of three measurements on each sample usually in the central region. We could not observe significant spectral difference among measurements on the same sample. Data were taken over two sets of measurements (three samples per measurement). 

\begin{figure}
\centering
\includegraphics[width=0.47\textwidth]{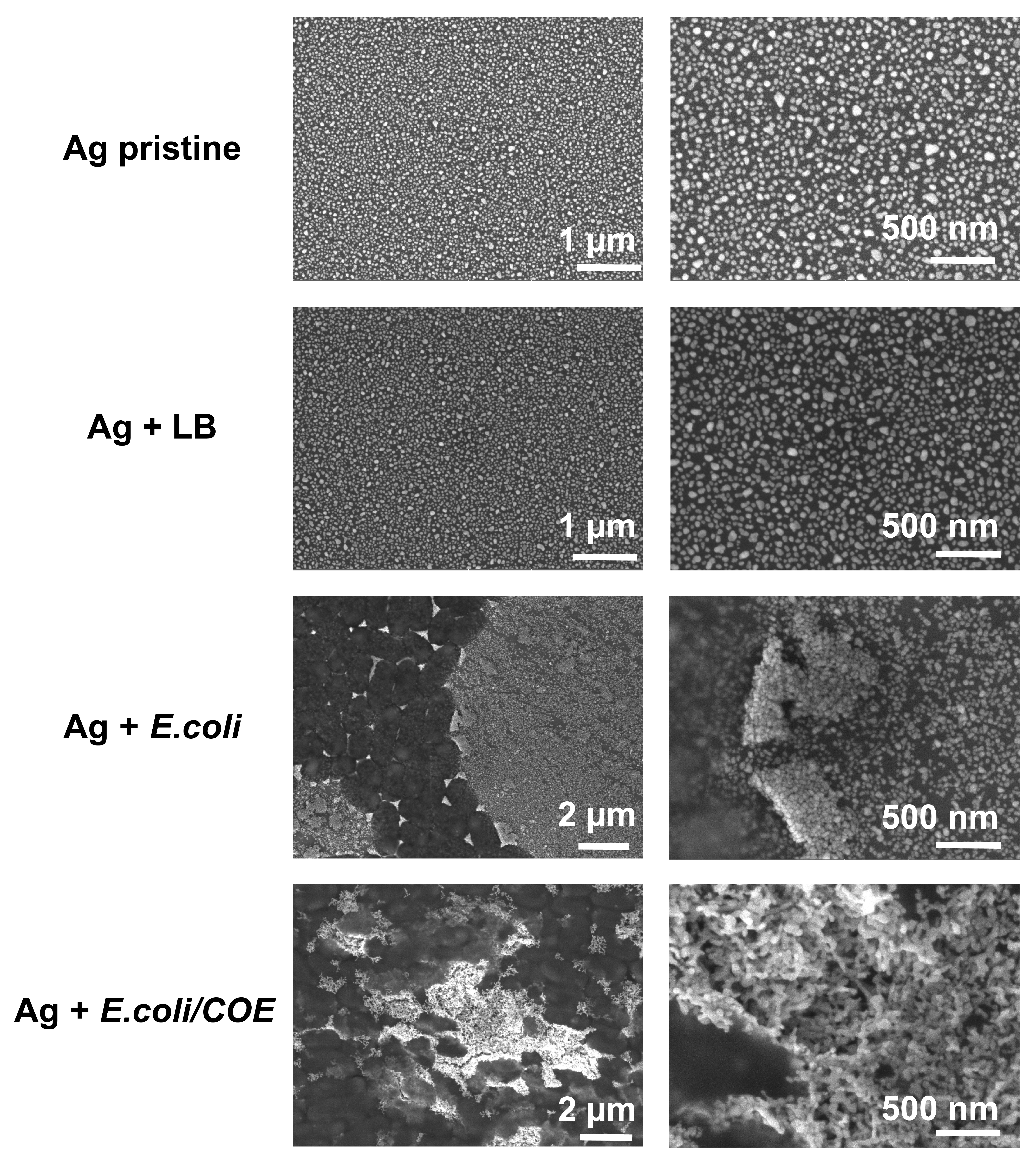}
\caption{SEM images of the silver NPs film in the pristine status, and after exposure to the culture medium LB (control measurement), \textit{E.coli} and \textit{E.coli}/COE. }
\label{fig:SEMmain}
\end{figure}

\begin{figure}
\centering
\includegraphics[width=0.47\textwidth]{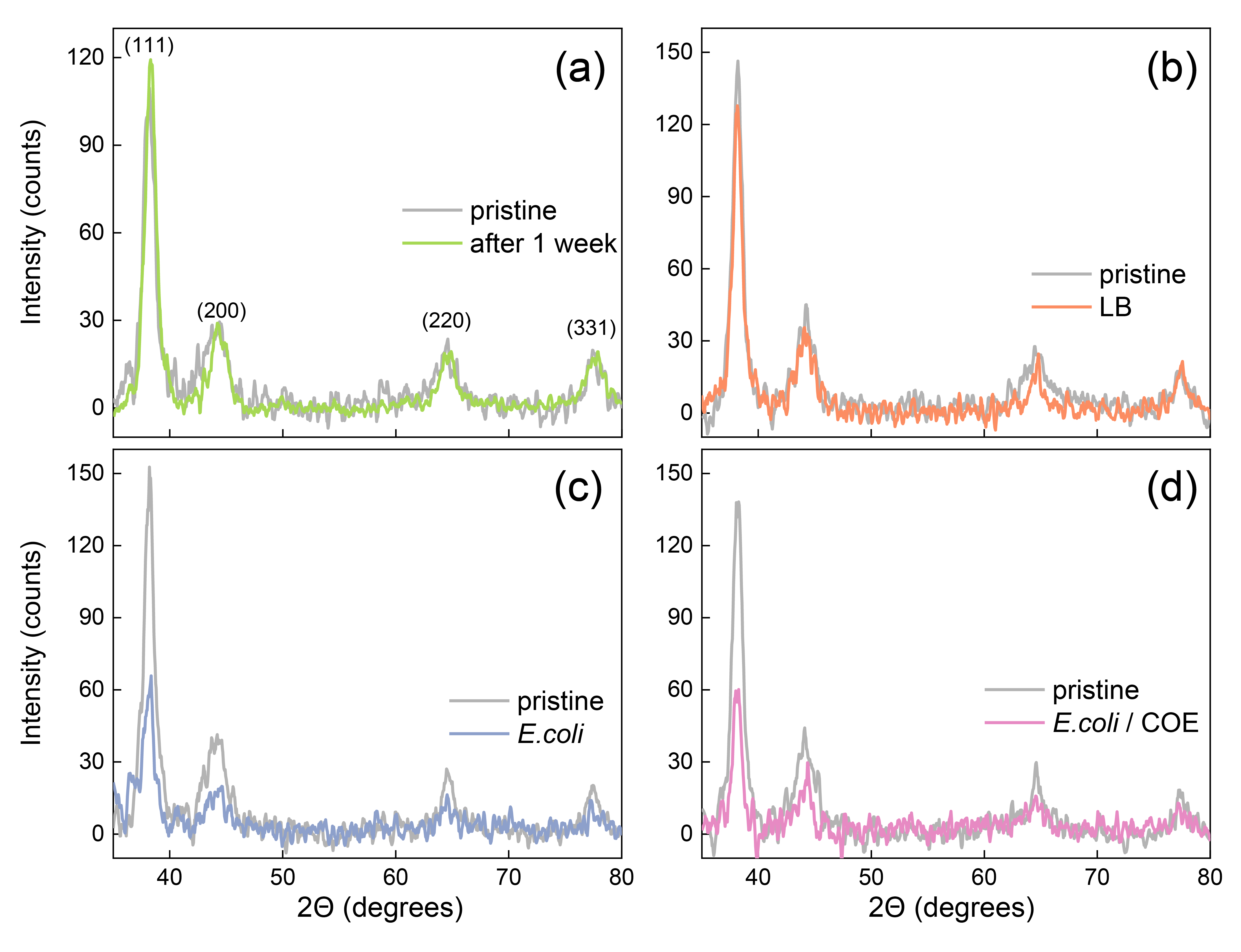}
\caption{(a) XRD pattern of silver, taken as deposited and after one week of storage. This control measurement was taken to account for the delay between silver deposition and actual XRD measurements (one week). Diffraction pattern of silver before and after exposure to LB (b), \textit{E.coli} (c) and \textit{E.coli}/COE (d). The measurements were taken by using a grazing-angle geometry ($\alpha$ = 0.5$^\circ$). }
\label{fig:XRD}
\end{figure}

The ultrafast pump-probe experiments described in the following sections proceeded in the following way. Ultrafast pulses were generated with a Ti:Sapphire chirped pulse amplified laser (Coherent Libra) with 2 kHz repetition rate, a center wavelength of 800 nm, and a pulse FWHM of around 150 fs. The fundamental pulse was split between a pump and probe path. In the pump path, second-harmonic generation was performed in a $\beta$-barium borate (BBO) crystal yielding 400 nm pump pulses with up to 500 $\mu$W average power at the sample; the fundamental was spectrally filtered with BG39 Schott glass filters. The pump is mechanically chopped at 1 kHz and delayed via a mechanical delay line out to a maximum of 1 ns pump-probe delay time. The pump is cross-polarized with the probe and focused down to a 150 $\mu$m diameter spot at the sample location with a reflective spherical mirror and spectrally rejected with an iris after the sample, as well as polarization-rejected with a polarization analyzer before the spectrometer slit. On the probe path, the fundamental is spatially filtered with an iris, attenuated, and focused into a translating calcium fluoride window for white light generation (WLG), and re-collimated by a reflective spherical mirror. The probe is focused onto the sample with < 100 $\mu$m diameter at a small angle, re-collimated with a lens and sent to a high-speed spectrometer (EB Streising). Measurement and analysis software includes Labview and Matlab 2020a. The data were collected over the course of one year during different experimental sessions. In order to compare spectra and kinetics measured with pump-probe from different sessions and to build statistical significance, the data were normalized for a chosen pump-probe delay time by dividing by the average magnitude of the signal. This method was observed to provide the highest reproducibility between data sets, allowing for the comparison of lineshapes even in the presence of fluctuations in absorbed pump power density due to observed variations in NP geometry and fill factors at different excitation spots. 

\subsection{SEM imaging and XRD}\label{subsec:SEM}

SEM measurements were carried out to investigate possible changes in shape and morphology of silver particles upon bacterial contamination (Figure \ref{fig:SEMmain}). The images of the pristine silver sample reveal the presence of nanoplates (NPs), with irregular shapes and sizes (average diameter = 18 nm, standard deviation = 6 nm, Figure S.1). While exposure to LB medium only does not lead to any appreciable effect on silver NPs, \textit{E.coli} cells cause dramatic changes in the surface morphology, with clear particle agglomeration in close proximity of the bacterial membrane and clusterization inside bacterial cells (see Figure S3). The cause of such an effect may be the enhanced surface reactivity due to the oxidative dissolution process \cite{Stark2011}. Furthermore, silver NPs seem to undergo a shape rounding-out and a size decrease within the aggregate, although particle fusion and severe aggregation hinder a more detailed and quantitative analysis over the size/shape distribution. Interestingly, addition of the COE molecule to the bacterial culture results in a more evident change in the silver NPs morphology. Specifically, while bacteria seem to be completely overwhelmed by the particles, the morphology changes from nanoplates typical of pristine samples to a worm-like complex network. Also in this case, such an intricate silver morphology hamper any further quantitative analysis of size and shape.

X-ray diffraction (XRD) analysis of the samples is presented in Figure \ref{fig:XRD}. The measurements were carried out by using a grazing-angle geometry (0.5$^\circ$), to maximize the signal coming from the silver surface. The diffraction pattern exhibits reflection peaks at 38.3$^\circ$, 44.3$^\circ$, 64.6$^\circ$ and 77.4$^\circ$, which are compatible with a face-centered cubic (fcc) packing \cite{Kora2013}. Here, we can see that while exposure to LB medium does not cause any modification of the XRD pattern, \textit{E.coli} and \textit{E.coli}/COE leads to a marked decrease of the fcc metallic silver peaks (60\% reduction). This strongly suggests the occurrence of an amorphization process originating from silver release due to oxidative dissolution \cite{Liu2010}.

\begin{figure}
\centering
\includegraphics[width=0.47\textwidth]{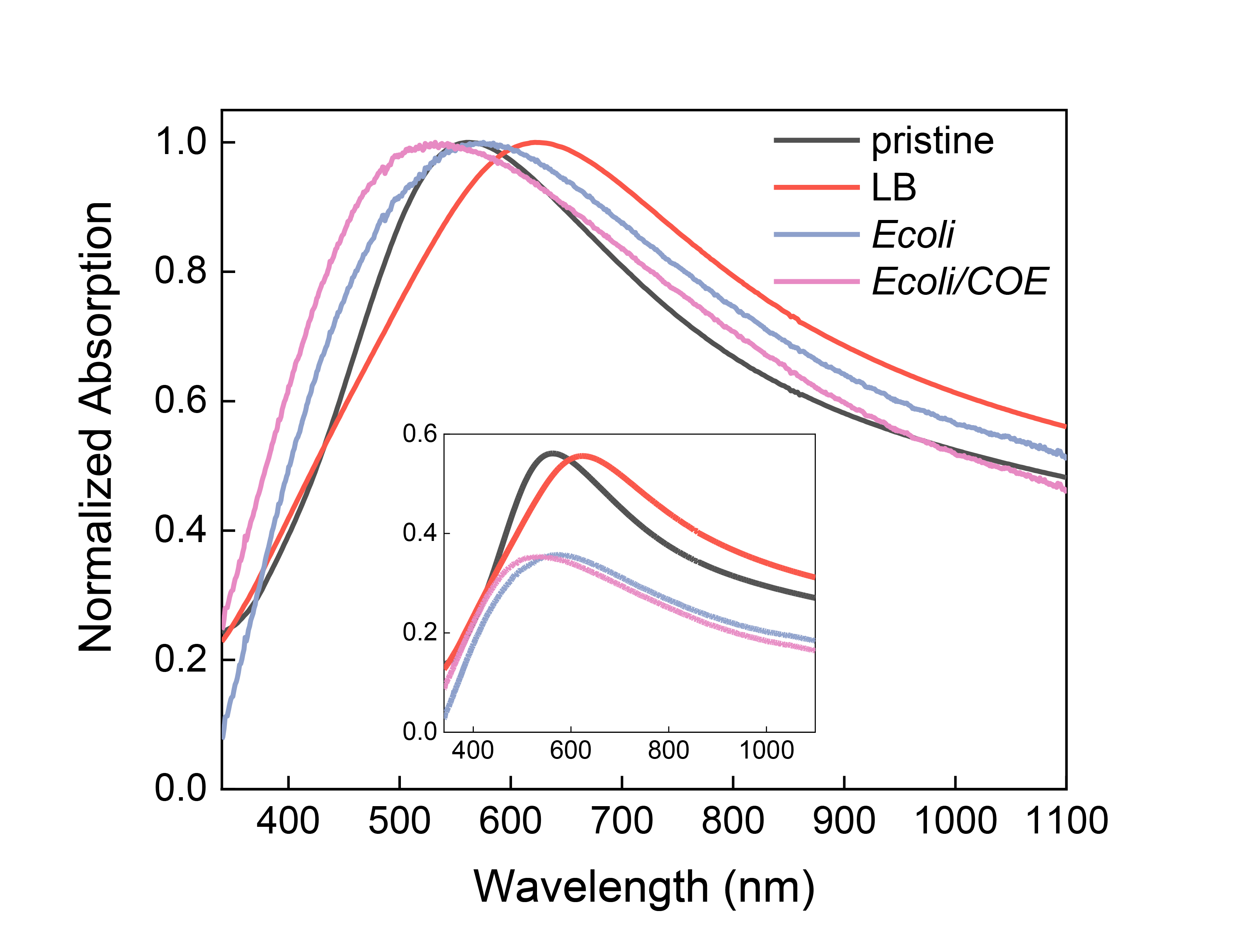}
\caption{Absorption spectra of silver NPs after exposure to LB (red line) \textit{E.coli} (blue line) and \textit{E.coli}/COE (purple line) normalized to maximum. The scattering of the \textit{E.coli} cells was subtracted from the transmission of silver exposed to \textit{E.coli}. The inset shows the non-normalized absorption spectra.}
\label{fig:linear}
\end{figure}

\subsection{Linear absorption}\label{subsec:linear}

The linear absorption spectra for the four samples are reported in Figure \ref{fig:linear}. The spectrum of pristine silver NPs exhibits a peak centered at 560 nm that broadens towards the near infrared (NIR). Usually, colloidal spherical silver nanoplates present a well-defined LSPR peak at 400 nm, while our silver nanoplates show a more complex spectrum due to the coexistence of different polarization modes in our high-aspect ratio particles. In particular, the 560 nm peak can be associated to the LSPR of nanoplates (average 18 nm) while the wide distribution of elongated and large particles contribute to the broad absorption in the NIR \cite{Metraux2005,Xue2007,Yi2013}. Although such a response is red-shifted by the culture medium only (LB) due to the increase in the environmental dielectric function, exposure to E.coli or to \textit{E.coli}/COE leads to different effects, namely: i. a severe broadening of the feature and a decrease of absorption intensity (Figure \ref{fig:linear} inset); ii. a blue-shift of the LSPR peak when compared with the silver/LB sample, which is enhanced by the COE permeability enhancer. The blue-shift and decrease of absorption intensity can be connected to the oxidative dissolution processes, especially if it is mediated by electron-donating nucleophiles (i.e. thiolates) \cite{Mogensen2014}, a process that eventually can lead to an increase of the charge carrier density. On the other hand, peaks broadening stems from the bacterial-driven enhancement of particles inhomogeneity, both in terms of shape and sizes.

\begin{figure}
    \includegraphics[width=0.35\textwidth]{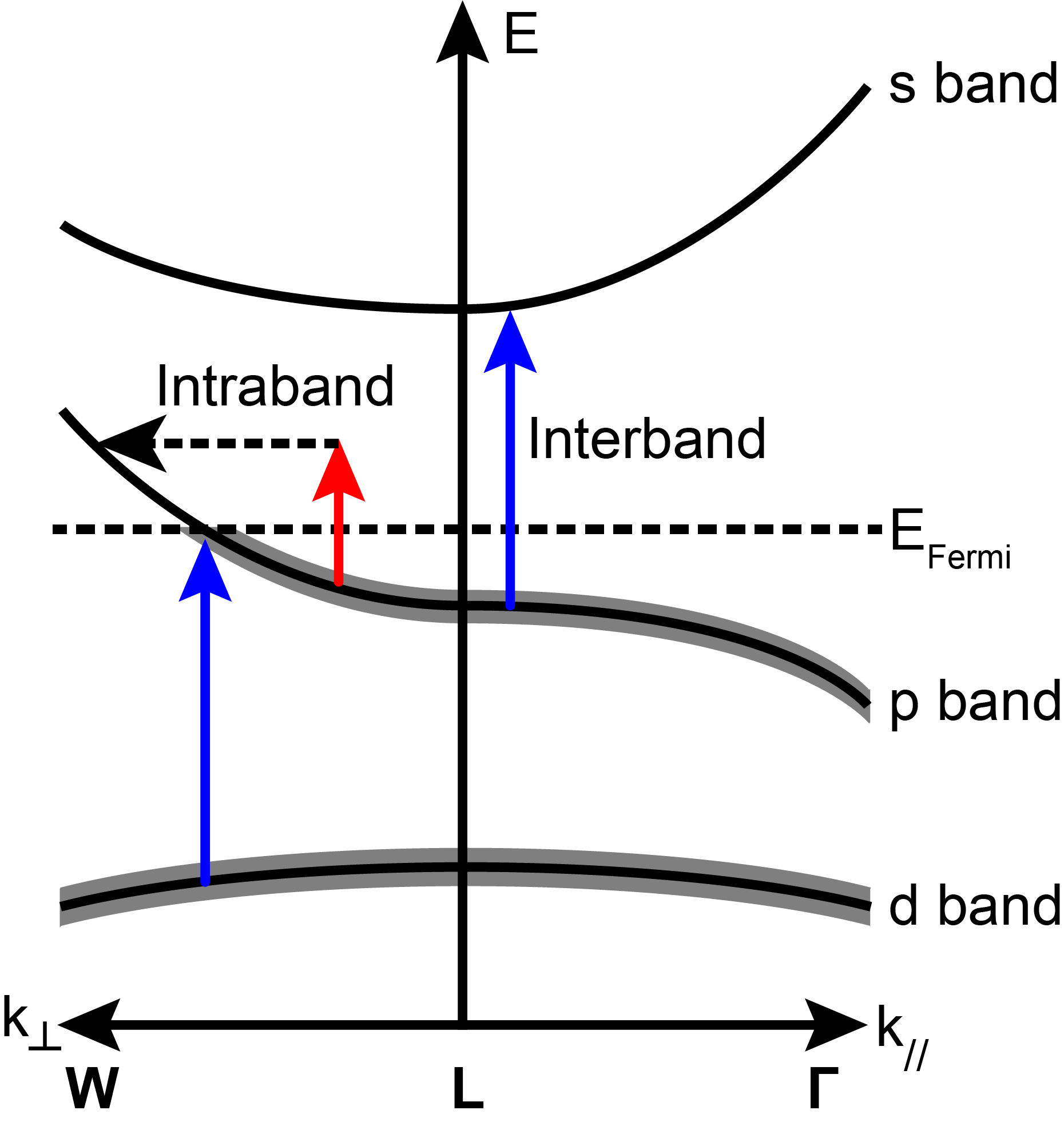}%
    \caption{Band structure of Ag in the vicinity of relevant optical transitions. Grey bands denote filled states below the Fermi energy. \label{fig:banddiagram}}
\end{figure}

\begin{figure*}
    \includegraphics[width=\textwidth]{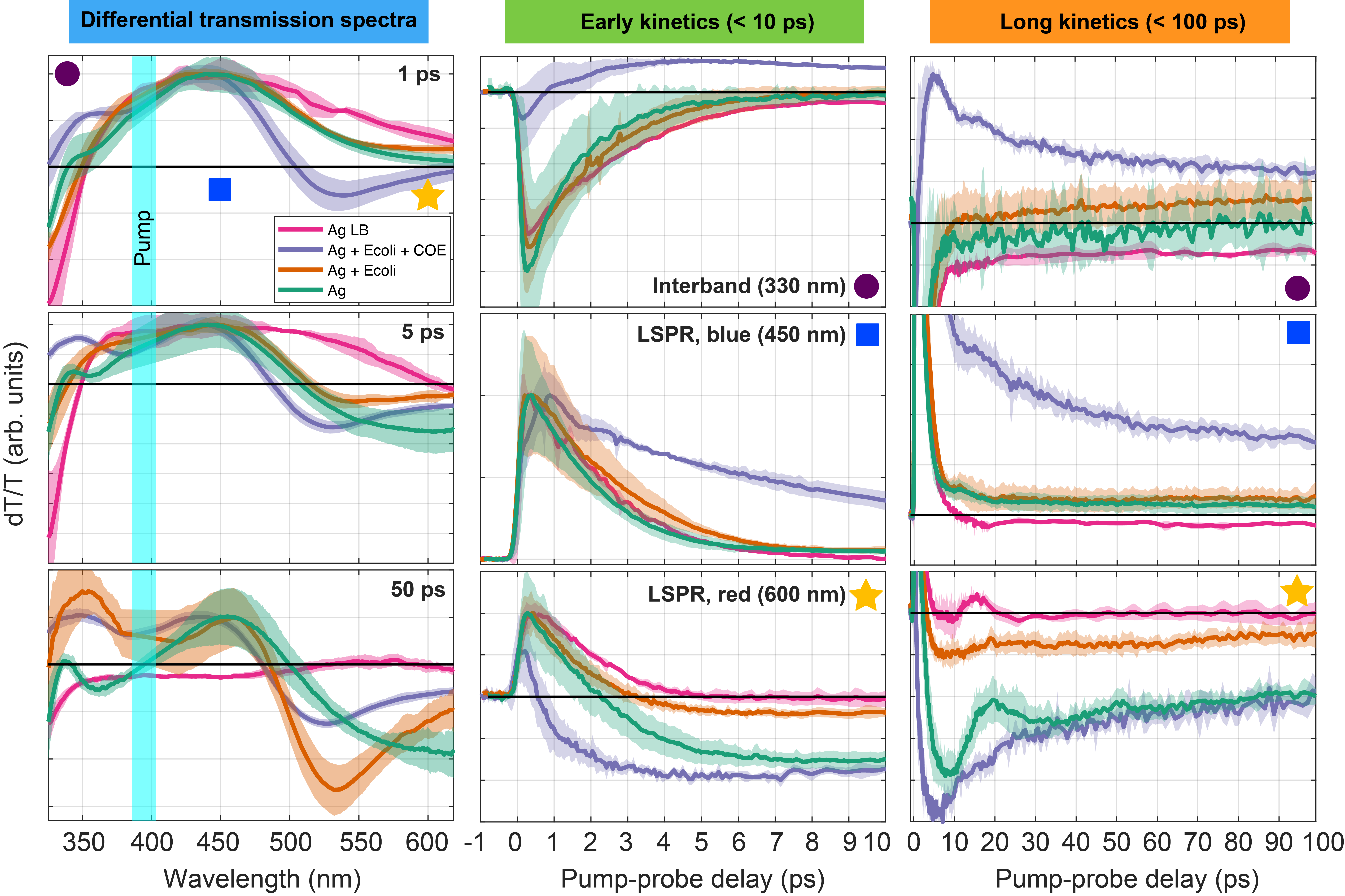}%
    \caption{Spectral and kinetic results of differential transmission (dT/T) ultrafast pump-probe experiments taken on Ag, Ag + \textit{E.coli}, Ag + \textit{E.coli} + COE, and Ag + LB samples. Each spectral and kinetic curve is the average over a large number of experimental runs, where each experimental result has been normalized by the average magnitude of the signal; confidence interval bands are also displayed. In the first column, the pump wavelength is indicated by a light blue band centered at 395 nm with 10 nm width. A circle, square, and star are used to indicate the corresponding spectral region for which the kinetics in the second and third columns are displayed, with the circle/square/star centered at 330, 450 and 600 nm, respectively. The second (third) column displays the kinetics at label and symbol-indicated wavelengths for sub 10 ps and sub 100 ps timescales, respectively. In the third column, some of the early time data have been cropped for increased clarity of long-time dynamics.\label{fig:averagedSpectra}}
\end{figure*}

\section{Pump probe (PP) studies of Ag NP thin films exposed to \textit{E.coli}/COE}\label{sec:PP}

The time-dependent non-linear optical response of Ag NPs can be described starting from two points (in addition to the plasmonic physics described in the supplementary information): 1. the band structure of Ag, described here, and 2. the two-temperature model, described in the supplementary information. First, the region of the band structure that is relevant at optical wavelengths \cite{Christensen1972, Rosei1974} is depicted in Figure \ref{fig:banddiagram}, with relevant optical pathways. Near the L point in the Brillouin zone, the d-, p-, and s- bands are considered, and the Fermi level overlaps with the p- band. Interband absorption, which occurs above 340 nm, takes place from the filled d-band to the portion of the p-band above the Fermi level, as well as from the portion of the p-band that is filled to the empty s-band. Intraband excitation, both via pump and probe in our experiments, occurs primarily in the p-band, but modifies the non-linear response of both interband and intraband excitation. Strong intraband pumping leads to a ``smearing'' of the Fermi level due to an increase in the electronic temperature \cite{Voisin2001, Groeneveld1995, Scotognella2013}. Lastly, the two-temperature model \cite{Kaganov1955,Fisica1994,DellaValle2012} describes the exchange of energy deposited by the ultrafast laser pulse into the plasmonic nanoparticle between the electrons, lattice, and surrounding environment.

\subsection{Pump-probe spectra}\label{subsec:spectra}

The results of the pump-probe (PP) experiments, performed on pristine Ag NP thin films and the bacteria-exposed samples are discussed in this section. We observed three common spectral features in all samples (Figure \ref{fig:averagedSpectra}): 1. Interband absorption at wavelengths shorter than 350 nm \cite{Johnson1972}, 2. A broad bleaching signal associated with the LSPR from 350-500 nm. 3. A pump-induced red-shift as indicated by a dispersive lineshape centered around 500 nm. 

The PP spectra, as observed via differential transmission (dT/T, see Methods section \ref{subsec:fabrication}) at selected pump probe delays (1, 5, 50 ps), early (<10 ps) and long (<100 ps) kinetics are displayed in Figure \ref{fig:averagedSpectra}. The average data and confidence intervals of the four sample types, pristine Ag NP thin films (Ag), Ag exposed to \textit{E.coli} (Ag+\textit{E.coli}), Ag exposed to \textit{E.coli} treated with COE (Ag+\textit{E.coli}/COE), and Ag including LB (Ag+LB) (analysis in supplementary information), are displayed. We note that control PP experiments were carried out on glass substrates with \textit{E.coli} deposited (without Ag NPs): no PP signal was observed.

First, modulation of the interband absorption (<350 nm) is observed. At times longer than 1 ps, this PIA signal apparently shifts to higher energies outside of the spectral region of our data acquisition. As the electronic temperature $T_e$ decreases due to electron-phonon coupling, the red-shift of the imaginary part of the interband dielectric function dies off. 

From approximately 350-500 nm (blue square in Figure \ref{fig:averagedSpectra}), a broad positive feature is observed in every sample; at times earlier than 1 ps, all samples also show a positive feature continuing out into the red (at least as far as 600 nm) except for Ag+\textit{E.coli}/COE. This positive feature is associated with the LSPR of the Ag NP thin film, which arises due to a pump-induced modification of the interband and Drude dielectric constants as connected to the electronic $T_e$ and lattice $T_L$ temperatures (see modeling section and Figure S4 in supplementary information). 

After 1-5 ps in the Ag, Ag+\textit{E.coli}, and Ag+\textit{E.coli}/COE samples, the positive features at wavelengths longer than 450 nm become negative, and a Lorentzian-derivative-like lineshape (dispersive) appears, with zero crossings between 465-530 nm, depending on the PP delay time. This zero-crossing is viewed as an indicator of the pump-induced spectral shift of the LSPR, which has been shown previously to approximately correspond to the LSPR energy \cite{Hamanaka1999}. The zero-crossing is observed to shift into the blue with increasing pump-probe delay time. After a few ps, $T_e$ approaches equilibrium with $T_L$, and both approach the temperature of the environment: as seen in the modeling results (see supplementary information), this cooling process leads to a relaxation at long times towards the static LSPR position. Significant differences are noted between sample types in this spectral region. At long times (50 ps), where the contributions to $\Delta\epsilon^{Ag}$ from the increase in the lattice temperature $T_L$ dominate, the zero-crossings of each sample type are observed at 500 (Ag), 485 (Ag+\textit{E.coli}), and 479 nm (Ag+\textit{E.coli}/COE). We interpret these differences in long-time zero-crossing positions as arising from differences in the LSPR energy in the absence of the pump; the PP measurements provide an apparently more sensitive measurement of this energy than linear absorption, which is hindered due to significant inhomogeneous broadening of the LSPR arising from NP geometrical variations, as well as larger probe spot size in linear absorption (1 cm) compared to pump-probe (100 $\mu m$). 

A simple increase in the environmental dielectric function will lead to a red-shift of the LSPR (Supplementary information): the opposite result, a blue-shift is observed here, which is enhanced in the COE-treated \textit{E.coli}-exposed Ag NP thin film sample. Thus, the exposure of the Ag NP thin film to \textit{E.coli} modifies $\epsilon^{Ag}$, potentially in the following ways: 1. An increase in the plasma frequency $\omega_p$ due to an increase in the free electron density $n_e$. This may occur due to electron donation from the nucleophiles of the \textit{E.coli}, likely followed by oxidative dissolution \cite{LeOuay2015,Mogensen2014}. In fact, Henglein \cite{Henglein1993} showed that nucleophile donation of electrons to Ag NPs results in a strong blue-shift of the LSPR via either an increase in the Fermi energy or a reduction in effective particle volume, leading to an increase in electron density. 2. A change in the NP geometry, due to a ``rounding-out'' of sharp hot spots \cite{Zhang2005}, as well as reduction in aspect ratio \cite{Xue2007}. Electrochemical dissolution of Ag leading to elimination of sharp features in NPs has been shown to blue-shift the LSPR. One indication of this effect is confirmed by examining the long wavelength features in both linear absorption (Figure \ref{fig:linear}) and the dT/T spectra at 50 ps. In Ag NP thin films, a broad negative band extends beyond 650 nm (outside of our acquisition range), but in the \textit{E.coli} treated samples, including \textit{E.coli}+COE, this long wavelength feature is significantly reduced, likely due to a significant blue-shift of the inhomogeneous distribution of nanoplates due to bacterial modification. We also report significantly larger linewidths in our samples (60-160 nm) compared to what has been reported in the literature (15-40 nm)\cite{Hamanaka1999, DelFatti2000,Link2005, Tsung2006, Lee2006, Hartland2011,Stoll2014}, which may be attributed to inhomogenenous broadening due to variations in NP aspect ratio.

\begin{figure*}
    \includegraphics[width=0.8\textwidth]{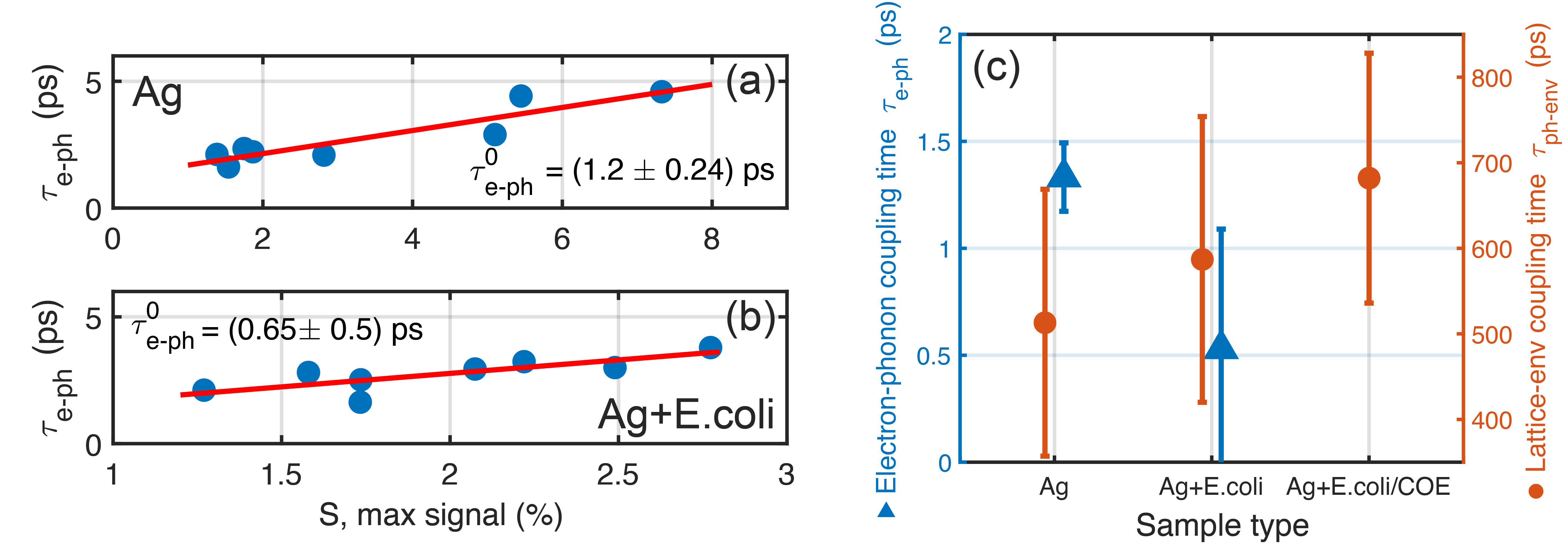}%
    \caption{Kinetic rates analysis of pump-probe results. Fit results for electron and lattice cooling parameters are shown, extracted by fitting kinetics data (shown in Figure \ref{fig:averagedSpectra}) as a function of maximum signals, instead of fluence. For the electron-phonon coupling curves (panels a and b), the resulting linear fit of the signal strength dependence is indicated in the plot. a, b: signal strength-dependent electron-phonon coupling times for Ag (a) and Ag+\textit{E.coli} (b). c: Compiled electron-phonon and lattice-environment coupling times for Ag, Ag+\textit{E.coli}, and Ag+\textit{E.coli}/COE (not included for $\tau_{e-ph}$); $\tau_{e-ph}$ indicated by blue triangles, $\tau_{ph-env}$ indicated by orange circles, with errorbars. 
    \label{fig:fluence}}
\end{figure*}

\subsection{Kinetics and fluence dependence}\label{subsec:kinetics}

Major differences between sample types are observed in the kinetics at early times (faster than 10 ps) and later times (between 10 and 100 ps). The kinetics are compared in the interband region (330 nm), and blue (450 nm) and red (600 nm) sides of the LSPR (second and third columns of Figure \ref{fig:averagedSpectra}). For PP delay times earlier than 10 ps, the differences in kinetics are most apparent on the red side of the LSPR (at 600 nm), where the aforementioned transition from bleaching to pump-induced absorption differs dramatically between sample types, but is reproducible over different experimental runs (see confidence interval bands). The kinetics for Ag+\textit{E.coli}/COE in this spectral range are clearly accelerated relative to the other samples. These very fast dynamics (faster than 500 fs) are common in all three spectral regions of Ag+\textit{E.coli}/COE. These rapid kinetics likely indicate a dramatic decrease in the electron-phonon coupling time compared to Ag or Ag+\textit{E.coli}; after 1 ps, the electron and lattice temperatures may nearly be in equilibrium, leaving only lattice cooling to the surrounding environment. This significant change in the electron-phonon coupling time is likely caused by the enhanced uptake of Ag ions into the \textit{E.coli} enhanced by treatment with COE \cite{Catania2016}. For later PP delay times (10<$\Delta t$<100 ps), the relative weighting of the slow decay for Ag + \textit{E.coli} + COE is large compared to the fast e-ph cooling kinetics, as evident in the interband region and the blue side of the LSPR. 

The significant decrease in the electron-phonon coupling time between Ag and Ag+\textit{E.coli}, especially the COE-treated sample, can be considered by examining the rate in terms of material parameters \cite{Qiu1992, Jiang2005}. The electron-phonon coupling rate $\gamma_{e-ph}$ can be expressed as

\begin{equation}
    \gamma_{e-ph} = \frac{\pi^2 m_e n_e c_s^2}{6 \tau_e(T_e) T_e}
\end{equation}

where $m_e,n_e,c_s,\tau_e$ are the electron effective mass, electron density, bulk speed of sound equal to $\sqrt{B/\rho_m}$ where B is the bulk modulus and $
\rho_m$ is the density of Ag, and electronic temperature-dependent electron scattering time, respectively. Thus, at least two potential effects may reasonably explain the observed increase in the electron-phonon coupling rate (decrease in coupling time). First, either electron donation to the Ag NPs from the \textit{E.coli} nucleophiles or positive Ag ion absorption into the \textit{E.coli} would lead to an increase in $n_e$. Second, a reduction in the time between electron scattering events may be caused by a reduction in Ag crystallinity (amorphization), as suggested by analysis of coherent oscillations (Section \ref{subsec:oscillations}). 

Electron-phonon and lattice-environmental coupling times can be more consistently compared by examining the fluence-dependence of the times. It is well-known \cite{Groeneveld1990, Link1999, Arbouet2003, Hartland2004} that for noble metal plasmonic systems that the two-temperature model discussed in this article leads to fluence-dependent electron-phonon coupling rate $\gamma_{e-ph}$, since the electronic heat capacity is a function of the electron temperature, which itself is a function of PP delay time and fluence (Supplementary information). It has been observed previously that $\tau_{e-ph}$ increases linearly with fluence, with zero-fluence results of $\tau_{e-ph} = $0.65 ps for Au NPs \cite{Hartland2004}, 0.67 ps for Ag thin films \cite{Groeneveld1990}, and 0.5-0.85 ps for Ag NPs ranging in radius from 3.2 to 30 nm \cite{Arbouet2003}.

A fluence-dependence analysis is performed on the Ag, Ag+\textit{E.coli} and Ag+\textit{E.coli}/COE samples. To determine the fast decay time (the electron-phonon coupling $1/\gamma_{e-ph}$), the early interband absorption kinetics at 330 nm were fit, while the later kinetics on the blue side of the LSPR (450 nm) were fit to determine the lattice-environment coupling. Initially no clear dependence of the fast decay times related to $\tau_{e-ph}$ was observed: we noticed during the experimental runs that for a given fixed input fluence the maximum signal varied for a given sample when different spots on the sample were excited. These variations are easily explained by the measured variations in particle size/geometry and fill factors at different spots on the sample (Section \ref{subsec:SEM}): the absorbed pump power, and therefore initial electron temperature, depends on all of these parameters (see Equation S.5). 

Instead, the decay times were studied as a function of maximum signal, which is a more direct indicator of the absorbed optical power, and still allows us to extract zero-fluence decay times (Figure \ref{fig:fluence}). Both $\tau_{e-ph}$ and $\tau_{ph-env}$ were examined as a function of maximum signal: only $\tau_{e-ph}$ showed a clear linear dependence. The electron-phonon coupling time in Ag+\textit{E.coli}/COE was not easily fit, since the kinetics were anomalously fast compared to Ag and Ag+\textit{E.coli}. $\tau_{e-ph}^0$ for Ag and Ag+\textit{E.coli} were determined to be equal to 1.2 $\pm$ 0.24 ps and 0.65 $\pm$ 0.5 ps, respectively. The electron-phonon coupling time has decreased in Ag+\textit{E.coli} compared to Ag; this trend is qualitatively consistent with Ag+\textit{E.coli}/COE, where the electron-phonon coupling time was too short to fit easily. Thus, it is reasonable that Ag release and uptake due to oxidative dissolution likely mediated by nucleophiles in the \textit{E.coli}, which increased with COE, caused this acceleration of electron-phonon cooling.

On the other hand, the lattice-environment coupling, which is very sensitive to the surrounding environment thermal conductivity \cite{Mohamed2001,Hu2002}, does not show a clear maximum signal dependence, but does show an increasing trend (in decay times) from Ag to Ag+\textit{E.coli} to Ag+\textit{E.coli}/COE, with $\tau_{ph-env}$ equal to 513 $\pm$ 156, 587 $\pm$ 167, and 682 $\pm$ 146 ps, respectively (Figure \ref{fig:fluence}c). Hu et. al \cite{Hu2002} showed that the decay time of thermal energy from the lattice into the environment is proportional to the surface area of the particle, and is independent of initial electron temperature, at least for an aqueous solution as surrounding environment: cooling times widely ranged from 10 to 380 ps for Au NPs ranging in radius from 5 to 50 nm. Mohamed et. al \cite{Mohamed2001} also showed clear dependence of the cooling rate of Au NPs embedded in gel matrices on the thermal conductivity and heat capacities of the gel, with cooling times ranging from 120 ps to 1.65 ns. Thus, even in the absence of \textit{E.coli}, the cooling time $\tau_{ph-env}$ of pristine Ag is on the high end of previous results; this likely results from the large surface area of the NPs. The clear increase in this cooling time may result from the change in the environment heat capacity/thermal conductivity of the \textit{E.coli} cells. This increase in the decay time of thermal energy to the environment is also considered in the context of the absence of coherent oscillations in Ag+\textit{E.coli} and Ag+\textit{E.coli}/COE, as discussed in Section \ref{subsec:oscillations}.

\subsection{Coherent oscillations}\label{subsec:oscillations}

Another apparent dynamic feature is the coherent oscillation with a period of between 15-30 ps present in Ag and Ag+LB; this phenomenon is not evident in the Ag+\textit{E.coli} and Ag+\textit{E.coli}/COE samples. Oscillations induced by ultrafast impulsive injection of acoustic phonons in both plasmonic and semiconductor nanocrystals have been studied intensely under variation of particle size and shape\cite{Nisoli1997,DelFatti1999,DelFatti2000,Qian2000,Hartland2002,Petrova2007}. For small initial electronic temperature changes, the change in the lattice temperature dominates in terms of contributions to the oscillations: as the lattice heats up, the particle expands, changing both the interband absorption edge and free electron density. 

In fact, since the oscillation period depends on the size and shape of the NPs, the damping time of the oscillation is often used as a measure of sample polydispersivity \cite{Hartland2002}. It is noted that in Ag and Ag + LB the oscillations are damped within one oscillation, indicating a large dispersion in particle geometries, as confirmed by SEM imaging (Section \ref{subsec:SEM}). The oscillator amplitude decay was determined to be approximately 12 ps via fitting; this number can be used to estimate the polydispersivity in NP size. Hartland \cite{Hartland2002} gives a relationship between the amplitude decay rate, oscillation period, and polydispersivity $\sigma_R$ (for Gaussian damping resulting from inhomogeneous broadening) for a collection of nanospheres as $\sigma_R = \bar{R}\bar{T}/\left(\sqrt{2} \pi \tau \right)$, where $\bar{R},\bar{T}$ are the average radii, oscillation period, and $\tau$ is the decay time: we arrive at $\sigma_R/\bar{R} =$ 0.48, compared to 6.4 nm$/$18 nm = 0.36 measured from SEM image analysis (Figure S.1). Thus, the explanation of oscillatory damping due to inhomogeneous broadening of the NP size/shape is qualitatively consistent with our observed variations in particle size, rather than loss of acoustic energy to the environment.

The absence of oscillations in the \textit{E.coli} exposed samples (including COE) is discussed below. First, we note for the \textit{E.coli} (and COE)-exposed Ag thin films that although Figure \ref{fig:averagedSpectra} displays the kinetics at 600 nm where the Ag sample has a strong pump-induced absorption (PIA) signal, oscillations are not observed at any wavelength in \textit{E.coli} (and COE). Second, as shown in the discussion of the fluence dependence, the decay times of lattice energy to the environment apparently increase in order from Ag to Ag+\textit{E.coli} to Ag+\textit{E.coli}/COE; thus, this factor can not explain the damping of oscillations. One possible explanation is a large increase in polydispersivity of the NP geometry inhomogeneous distribution. As discussed earlier, it has been shown that electrochemical dissolution can lead to a ``rounding-out'' effect of NPs, eliminating hot spots and leading to a blue-shift of the LSPR \cite{Zhang2005}. We posit that this dissolution effect may occur due to bacterial absorption of Ag ions into the cell. A possible explanation is modification of the Ag NP surface due to oxidative dissolution by the \textit{E.coli}\cite{LeOuay2015,Mogensen2014,Henglein1993}. This surface modification may lead to an amorphization of the Ag NP, which is expected to modify the acoustic phonon modes, potentially leading to destruction of the coherent oscillatory mode. Amorphization of the Ag NPs induced by bacterial action is indicated by the 60\% reduction in XRD scattering intensity (Figure \ref{fig:XRD}). Another potential explanation may arise from the aggregation of Ag NPs into islands, leading to modification of the Ag NP at the new interfaces; disorder arising from this interfacial disruption may wash out the oscillatory modes.

\section{Conclusions and future directions}\label{sec:conclusions}
We have presented a detailed analysis of the plasmon response of silver NPs upon exposure to \textit{E.coli}. Although silver has been extensively studied for its antibacterial properties, the possible effects of bacterial contamination on its optical features have remained largely unexplored. Here, we combine morphological, structural, and optical data with ultrafast pump-probe characterizations to shed light into these aspects. We found that silver NPs undergo dramatic changes both in terms of particle shape/size and crystallinity, most likely owing to silver oxidative dissolution. In particular, we observe clear agglomeration of silver NPs at the close proximity of the membrane and clusterization inside \textit{E.coli} cells. Within these aggregates, we can qualitatively note a decrease of particles size and a rounding-out of nanoplate edges. Furthermore, the addition of a membrane permeabilizer (COE) to the bacterial culture leads to a clear enhancement of these effects.

Summarizing, the exposure of the Ag NPs to bacteria induces the following changes in the pump-probe response. First, the electron-phonon coupling time decreases from Ag to Ag+\textit{E.coli} to Ag+\textit{E.coli}/COE, explained by either an increase in the free electron density or amorphization of the Ag NPs leading to faster electron-electron scattering. Second, the lattice-environment coupling time increases in the same order. Third, coherent oscillations that are observed in Ag are no longer observed in the \textit{E.coli}-exposed samples. The lack of oscillations may be explained again by either amorphization of the NPs at the surface due to bacterial exposure, or an increase in polydispersivity of the NP geometries. Last, a static blue-shift of the dispersive feature associated with the LSPR is observed in Ag+\textit{E.coli} and Ag+\textit{E.coli}/COE compared to Ag, as revealed by the enhanced sensitivity of pump-probe compared to linear absorption. All of the spectroscopic results may be explained by either an increase in the free electron density via nucleophile-mediated oxidative dissolution, NP geometrical changes (size reduction, rounding-out), and particle amorphization, all driven by the silver/bacteria interaction.

In the next experiments, we aim to explore in more details the bacteria-induced morphological/structural changes within silver NPs aggregates and the chemical pathways stemming from such an interaction, by means of a more in-depth microscopy (i.e. via atomic force and transmission electron microscopies) and spectroscopic analysis (i.e. by using vibrational spectroscopies). In both cases, the main experimental difficulty resides on the complex environment offered by the biological setting, which leads to severe particle aggregation and to a plethora of electrochemical reactions that might occur between silver and bacteria. Nevertheless, we reckon that a lack of fundamental physical understanding hampers the prediction of the material behavior upon contamination with bacteria. Therefore, this study illustrates innovative research opportunities in the field of biophysics.

\section{Acknowledgments}

This work has been supported by Fondazione Cariplo, Grant No. 2018-0979. F.S. and A.M.R. thank the European Research Council (ERC) under the European Union’s Horizon 2020 research and innovation programme (Grant Agreement No. [816313]). The following article has been submitted to Applied Physics Reviews. After it is published, it will be found at \url{https://publishing.aip.org/resources/librarians/products/journals/}.

%

\end{document}


\preprint{AIP-AgBacteria}

\title[Supplementary information]{Supplementary information for: The Impact of Bacteria Exposure on the Plasmonic Response of Silver Nanostructured Surfaces}

\author{Giuseppe M. Patern\`{o}}
\thanks{These authors contributed equally to this work.}
\affiliation{ 
Center for Nano Science and Technology, Istituto Italiano di Tecnologia (IIT), Via Pascoli 10, 20133, Milano, Italy
}
\author{Aaron M. Ross}%
\thanks{These authors contributed equally to this work.}
\affiliation{ 
Physics Department, Politecnico di Milano, Piazza L. da Vinci 32, 20133 Milano, Italy
}%
\author{Silvia M. Pietralunga}
\affiliation{ 
Center for Nano Science and Technology, Istituto Italiano di Tecnologia (IIT), Via Pascoli 10, 20133, Milano, Italy
}
\affiliation{ 
Institute for Photonics and Nanotechnologies (IFN), Consiglio Nazionale delle Ricerche (CNR), Piazza L. da Vinci 32, 20133 Milano, Italy
}
\author{Simone Normani}
\affiliation{ 
Center for Nano Science and Technology, Istituto Italiano di Tecnologia (IIT), Via Pascoli 10, 20133, Milano, Italy
}
\author{Nicholas Dalla Vedova}
\affiliation{ 
Center for Nano Science and Technology, Istituto Italiano di Tecnologia (IIT), Via Pascoli 10, 20133, Milano, Italy
}
\author{Jakkarin Limwongyut}
\affiliation{ 
Center for Polymers and Organic Solids, Department of Chemistry and Biochemistry, University of California, Santa Barbara, CA 93106, USA
}
\author{Gaia Bondelli}
\affiliation{ 
Center for Nano Science and Technology, Istituto Italiano di Tecnologia (IIT), Via Pascoli 10, 20133, Milano, Italy
}
\affiliation{ 
Physics Department, Politecnico di Milano, Piazza L. da Vinci 32, 20133 Milano, Italy
}%
\author{Liliana Moscardi}
\affiliation{ 
Center for Nano Science and Technology, Istituto Italiano di Tecnologia (IIT), Via Pascoli 10, 20133, Milano, Italy
}
\affiliation{ 
Physics Department, Politecnico di Milano, Piazza L. da Vinci 32, 20133 Milano, Italy
}%
\author{Guillermo C. Bazan}
\affiliation{ 
Center for Polymers and Organic Solids, Department of Chemistry and Biochemistry, University of California, Santa Barbara, CA 93106, USA
}
\affiliation{ 
School of Chemical and Biomedical Engineering Nanyang Technological University Singapore 639798, Singapore
}
\affiliation{ 
Singapore Centre on Environmental Life Sciences Engineering Nanyang Technological University Singapore 639798, Singapore
}
\author{Francesco Scotognella}
\email[Authors to whom correspondence should be addressed: ]{francesco.scotognella@polimi.it, guglielmo.lanzani@iit.it}
\affiliation{ 
Center for Nano Science and Technology, Istituto Italiano di Tecnologia (IIT), Via Pascoli 10, 20133, Milano, Italy
}
\affiliation{ 
Physics Department, Politecnico di Milano, Piazza L. da Vinci 32, 20133 Milano, Italy
}%
\author{Guglielmo Lanzani}
\email[Authors to whom correspondence should be addressed: ]{francesco.scotognella@polimi.it, guglielmo.lanzani@iit.it}
\affiliation{ 
Center for Nano Science and Technology, Istituto Italiano di Tecnologia (IIT), Via Pascoli 10, 20133, Milano, Italy
}
\affiliation{ 
Physics Department, Politecnico di Milano, Piazza L. da Vinci 32, 20133 Milano, Italy
}%

\date{\today}
\maketitle

\section{Ag pristine SEM image analysis}

\begin{figure}[h!]
\centering
\includegraphics[width=1\textwidth]{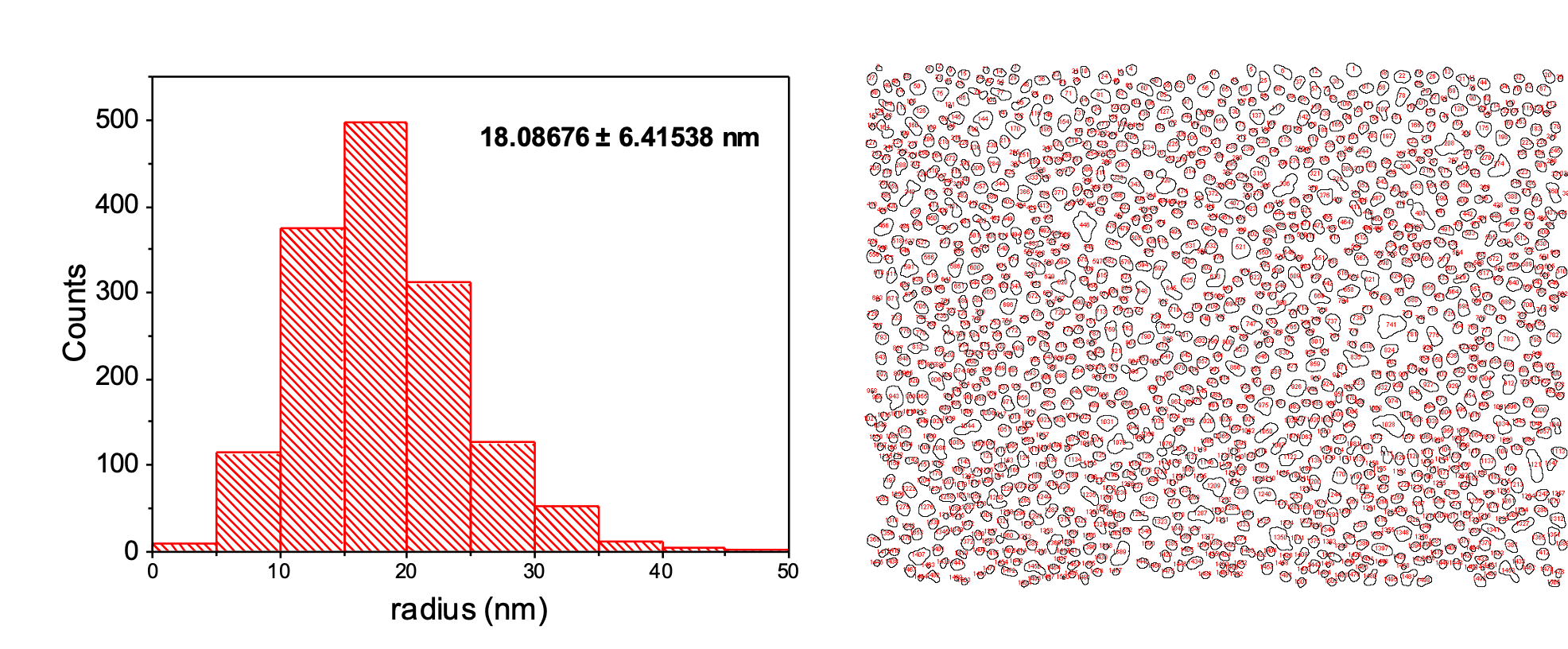}
\caption{(a) Histogram of size distribution of pristine silver NPs obtained via thermal evaporation on glass substrates. (b) Example of size and shape calculation of silver NPs. SEM images were analysed by using the ImageJ software.}
\label{fig:SEMsupp}
\end{figure}

\section{COE molecular structure}

\begin{figure}[h!]
\centering
\includegraphics[width=0.8\textwidth]{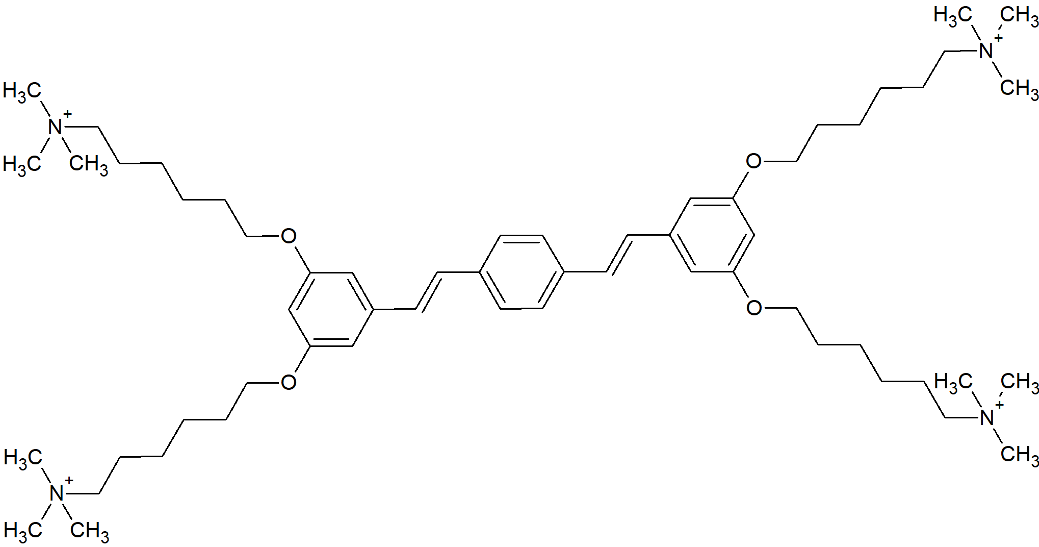}
\caption{Chemical structure of conjugated oligoelectrolyte used in this work (COE2-3C).}
\label{fig:COE}
\end{figure}

\section{Additional SEM silver/\textit{E.coli} images}

\begin{figure}[h!]
\centering
\includegraphics[width=\textwidth]{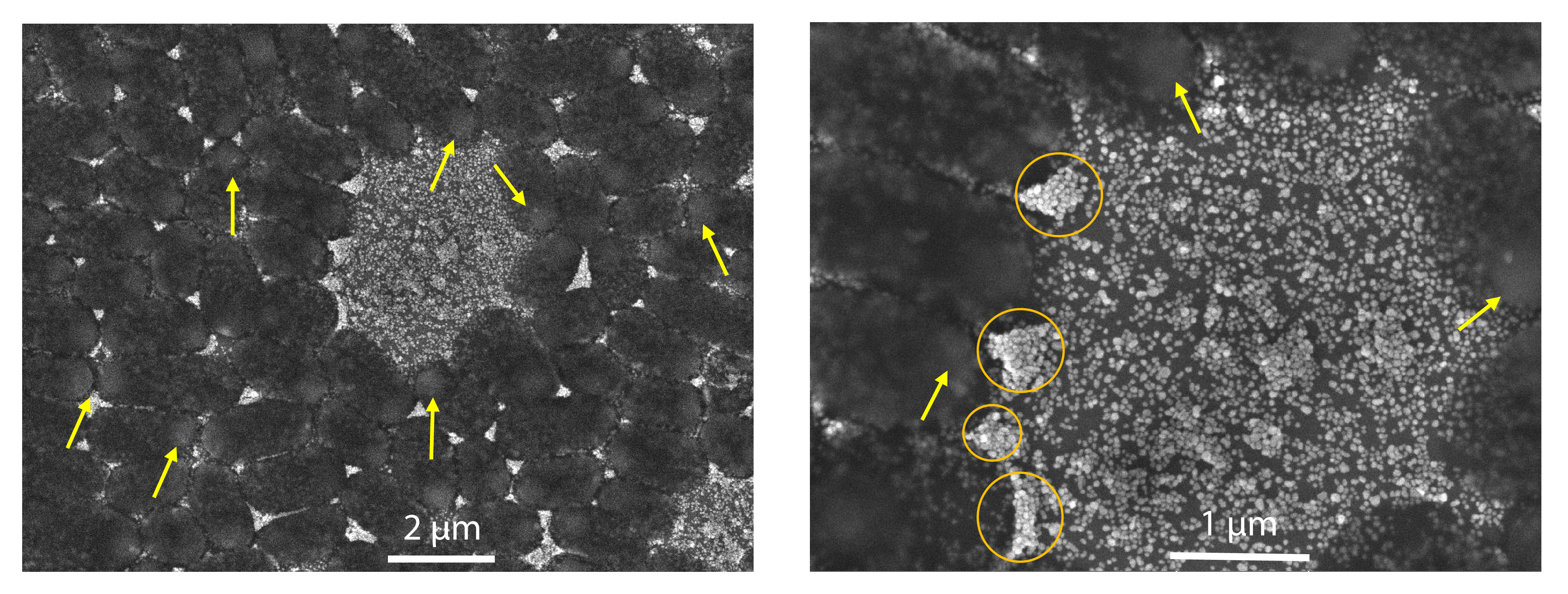}
\caption{SEM images of the silver NPs surface after contamination with E.coli cells. Yellow arrows highlight the presence of Ag clusters inside the cells, while orange circles indicate the Ag aggregates at the close proximity of the cell membrane.}
\label{fig:S3}
\end{figure}

\section{Spectral shifts of LSPR}
To better understand the static shifts in apparent LSPR energies between the Ag NP thin films and those treated by E.coli and E.coli+COE, a simple model is examined. A small Ag nanosphere (neglecting scattering) for which only the Drude free electron model is included (neglecting interband contributions), is considered. The LSPR condition, ie. that $\text{Re}\left[\epsilon_{Drude}\right] + 2 \epsilon_m = 0$, results in 

\begin{equation}
    \omega_{LSPR} = \sqrt{\frac{\omega_p^2}{1+2 \epsilon_m}-\gamma^2} \sim \frac{\omega_p}{\sqrt{1+2 \epsilon_m}}
\end{equation}

where $\omega_p, \gamma$ are the plasma frequency and Drude scattering rate \cite{Maier2007}. Thus, changes in either the plasma frequency, which may result for electron concentration changes (volume changes), or static changes in the surrounding dielectric constant, will result in the following shift in LSPR energy

\begin{equation}\label{eq:deltaEps}
    \Delta \omega_{LSPR} = \frac{1}{\sqrt{1+2 \epsilon_m}} \Delta \omega_p - \frac{\omega_p}{(1+2\epsilon_m)^{3/2}}\Delta \epsilon_m
\end{equation}

Thus, it is seen that an increase in the surrounding dielectric constant (ie. from air to water/E.coli (n = 1.33)) will result in a red-shift of the LSPR, while an increase in the plasma frequency, potentially through a donation of electrons into the NP, will result in a blue-shift of the LSPR. These results will be taken into consideration during interpretation of the PP data. The spectral shifts of the LSPR may also be explained by a reduction in the sharpness/hot spot features of the NP under oxidative dissolution \cite{Zhang2005}; a "rounding-out" out of these features will result in a blue shift of the LSPR.

\section{Pump-probe results for Ag + LB}
One other sample type was investigated in which the Ag NP thin film sample was contacted with the LB (lysogeny broth) agar growth plate without E.coli. As described earlier in the article, the growth medium is a blend of proteins, vitamins, trace elements, minerals, and especially NaCl, providing Na+ ions for cell function. The PP spectra and kinetics at early times (< 1 ps) resemble that of the pristine Ag NP thin films, with a broad positive feature from 360-650 nm, and interband pump-induced-absorption at shorter wavelengths (330 nm). While it is apparent from the interband absorption that the kinetics seem to be governed by the two-temperature model, the plasmonic features centered around 460 nm in the pristine Ag thin films are completely absent at later times, with only the electronic temperature kinetics evident at 450 and 600 nm. A small oscillatory feature with a similar period to that of the pristine Ag NP samples is observed, but the amplitude of the signal dies off within 30 ps. Thus, we believe that while the electron-phonon coupling and lattice-environment coupling do not seem to be modified significantly in Ag+LB compared to pristine Ag NPs, that the plasmonic features obvious in the other three sample types seem to be ``washed out'' in this sample. One potential explanation is that the environmental dielectric environment may be more heterogeneous in Ag+LB due to the large mixture of analytes in the LB medium that the Ag NP local electric field environment interacts with, thereby leading to significant inhomogeneous broadening by modifying the LSPR of each plasmonic NP. These analytes, or nutrients for the E.coli, are mostly consumed during the process of cell growth, and probably do not play a role in the Ag+E.coli samples.

\section{Modeling and interpretations}

As evidenced by the linear absorption and SEM image analysis results reported in Sections III.B, III.C, and the supplementary information, the control sample under study consists of Ag nanoplates (NP) with a lateral semi-axis (radius) distribution centered at 18 nm for the pristine Ag samples, with a standard deviation of at least 6.4 nm (Figure S1). The thickness of the NPs was determined by SEM analysis to be 8 $\pm$ 0.5 nm; thus, these NPs exhibit large aspect ratios, and also large variations in the aspect ratio. It has been shown previously that plasmonic nanoplate aspect ratio is the dominant determining factor of the localized surface plasmon resonance (LSPR) energy compared to the overall size of the particle \cite{Maier2007,Kreibig1995}. These large variations in aspect ratio, size, and thickness lead to significant inhomogeneous broadening of the LSPR resonance. Full Mie theory calculations \cite{Hu2008} predict linewidths between 100-250 meV (16-40 nm at 450 nm) for Ag nanospheres in air with radii varying between 5-35 nm; here, both linear absorption (Section III.C) and PP results (Section IV) show much broader linewidths, at least as large as 70 nm, indicating that geometrical variation in particle size is leading to significant inhomogeneous broadening. 

The optical absorption, as well as the transient optical absorption of the probe induced by intraband pump absorption, of the Ag NP system can be described by the polarizability $\alpha$ of the NP, leading to the extinction $\sigma_{ext}$, scattering $\sigma_{sca}$, and absorption $\sigma_{abs}$ cross-sections, which are given by \cite{Maier2007}

\begin{equation}\label{eq:scattering}
    \sigma_{ext} = \sigma_{sca} + \sigma_{abs} = \frac{k^4}{6\pi} |\alpha|^2 + k \text{Im}\left[\alpha\right]
\end{equation}

where k is the wave-vector of the pump/probe in the surrounding medium; the polarizability is determined from the boundary conditions dictated by Maxwell's equations, and include the particle geometry, Ag and surrounding media dielectric constants \cite{Jackson1975}.

The differential transmission that is measured in the following pump-probe experiments is related to the pump-probe delay-dependent differential extinction cross-section $\Delta \sigma_{ext}$, given by 

\begin{equation}
    \frac{dT}{T} = e^{-\Delta \sigma_{ext} N_p L}-1 \sim -\Delta \sigma_{ext} N_p L
\end{equation}

where $N_p$ is the particle concentration, and L is the sample path length.

To further confirm our understanding of the Ag and E.coli-exposed NP films, the system was modeled using the following procedure. First, for a given pump fluence and pulse width of 150 fs, the differential equations of the two-temperature model (Eqs \ref{eq:twotemp}) are solved numerically, resulting in $T_e(t), T_L(t)$. 

The two temperature model has long been utilized to treat the increase in the electronic $T_e(t)$ and lattice $T_L(t)$ temperatures due to ultrafast excitation \cite{Kaganov1955, Fisica1994,DellaValle2012}. At very early times, sub 100 fs, an ultrafast pulse leads to coherent excitation of oscillations of the the free electrons in Ag, as enhanced by the plasmonic resonance; this coherent excitation dies off via Landau damping within a few hundred fs \cite{Khurgin2017, DellaValle2012}, after which point the non-thermal electron distribution generated by the optical excitation thermalizes. The extended two temperature model \cite{Fisica1994, DellaValle2012, Heilpern2018} treats the decay of this non-thermal energy into both the thermal electronic distribution described by the Fermi-Dirac distribution, as well as directly to the lattice; this thermalization has been shown to be completed within between 0.5-2 ps \cite{Fisica1994}, and thus competes with electron-phonon coupling, which distributes energy between the hot (described by a temperature) thermal electron distribution and the lattice on timescales between 0.5 to 10 ps. Finally, the lattice of the NP cools to its environmental surroundings typically on much longer timescales than the electron-phonon coupling times, between 5-1000 ps. These timescales are discussed in more detail in Section IV.B, and are very sensitive indicators of both the local environment experienced by the plasmonic NP, as well as material changes induced on the NP by its environment, for instance due to bacterial action. 

The details of the two temperature model, including the differential equations, can be found in the supplementary information. There, it is shown that the electronic temperature immediately after ultrafast excitation is given by

\begin{equation}\label{eq:tempAfter}
    T_{e,\text{after}} = T_{e0}\sqrt{1+\frac{1}{\gamma_e T_{e0}^2}\frac{AF}{fL}}
\end{equation}

where $\gamma_e T_e$ is the electronic temperature-dependent electronic specific heat, A is the pump absorption, F is the fluence, f is the fill factor, and L is the absorptive path length. For average pump powers of 10-200 $\mu$W at 1 kHz repetition rate and a spot diameter of 150 $\mu$m (Section III.A), hot electron distribution temperatures of between 600-3000 K are generated. However, it is evident from Equation \ref{eq:tempAfter} that the change in electronic temperature after the pump pulse, and thus the maximum pump-probe signal, depends on parameters that vary considerably in each sample and between sample types, including the NP fill factor $f$ and the pump absorption A. 

The differential equations used in this article governing the basic two-temperature model without considering the non-thermal electronic distribution are given by \cite{Scotognella2011}

\begin{equation}\label{eq:twotemp}
\begin{split}
\gamma_e T_e\frac{dT_e}{dt} = &-\gamma_{e-ph}(T_e(t)-T_L(t)) + P(t), \\
C_L \frac{dT_L}{dt} = &+\gamma_{e-ph}(T_e(t)-T_L(t)) \\ 
&- \gamma_{ph-env}(T_L(t)-T_0)
\end{split}
\end{equation}

where $\gamma_e T_e$ is the electronic temperature-dependent electronic specific heat, $\gamma_{e-ph}$ is the electron-phonon coupling rate, $\gamma_{ph-env}$ is the lattice-environment coupling, $C_L$ is the lattice specific heat, and P(t) is the absorbed pump power density. The electronic specific heat is much smaller than the lattice specific heat \cite{DellaValle2013}, and thus the hot electron distribution can be heated up to > 1000 K for input fluences of greater than 0.1 mJ cm$^{-2}$. Specifically, assuming that the pulse duration is much shorter than the electron-phonon thermalization time, the absorbed power density is given by

\begin{equation}
    P(t) = \sqrt{\frac{4ln2}{\pi}}\frac{AF}{fL\tau}\text{exp}\left(-\frac{4ln2}{\tau^2}t^2\right)
\end{equation}

where A is the pump absorption, F is the fluence, f is the fill factor, L is the absorptive path length, and $\tau$ is the pulse FWHM. The expression in the main article for the electronic temperature after ultrafast pulse excitation can be solved for by setting $\gamma_{e-ph}$ = 0 and solving the first differential equation in the two-temperature model. This is the same as assuming 1/$\gamma_{e-ph}$ is much longer than the pulse width $\tau$.

Then, the method first developed by Rosei \cite{Rosei1974} is utilized: this method connects the time-dependent electronic temperature to changes in the imaginary part of the interband dielectric function of Ag $\epsilon_{Im}^{IB}$, which includes transitions from the d to p bands (in the vicinity of the Fermi level), and p to s bands. The change in the real part of the dielectric function $\epsilon_{Re}^{IB}$ is then determined via the Kramers-Kronig relations: this approach is acceptable since the change in the imaginary part of the interband dielectric function is well-localized around the interband edge; however, the Kramers-Kronig transformation leads to a long tail in the change in the real part far off into the near-IR spectral region. We note that this treatment requires fitting the imaginary part of the total dielectric function \cite{Johnson1972} with both the Drude and interband terms; the resulting fit parameters, along with other numerical parameters used in the following modeling are listed here: $E_{dp} = 3.5186$eV, $E_{s}$ = 4.0862 eV, $E_f$ = 0.31 eV, A$_{dp}$ = 1.8738, A$_{ps}$ = 2.2545, E$_p$ = 9.1 eV, $\gamma$ = 0.031 fs$^{-1}$, m$_{s\parallel}$ = 0.128, m$_{s\bot}$ = 5.16, m$_{p\parallel}$ = 0.172, m$_{p\bot}$ = 0.32, m$_{d\parallel}$= 2.075, m$_{d\bot}$ = 2.58, a = 0.402 nm, A$_{tot}$ = 10$^{48}$, $\gamma_e$ = electron heat capacity = 65 J $m^{-3}K^{-2}$, C$_{lat}$ = 2.4x10$^6$ J $m^{-3}K^{-1}$, G$_{e-ph}^0$ = 3x10$^{16}$ W $m^{-3}K^{-1}$, G$_{ph-env}^0$ = 1-40 x (5x10$^{14}$)W $m^{-3}K^{-1}$, absorbance = 0.1, fluence = 0.1-1 mJ cm$^{-2}$, fill factor = 0.31.

The electronic temperature $T_e(t)$ also affects all of the electronic scattering processes, ie. $\Delta \gamma = \Delta \gamma_{e-e} + \Delta \Gamma_{e-ph} +  \Delta \gamma_{e-s}$ \cite{Stoll2014}, through changes in the electron-electron scattering rate, electron-phonon scattering rate, and the surface-scattering rate which is relevant for plasmonic NPs with feature sizes comparable to the mean free path in Ag \cite{Stoll2014}. The change in the electron-electron scattering rate for changes in the electronic temperature below 2000 K is given by \cite{Gurzhi1959,Voisin2001,Stoll2014}

\begin{equation}
\frac{\Delta\gamma_{e-e}}{\gamma_{e-e}} \cong \left(\frac{2\pi k_B}{\hbar \omega}\right)^2 \left(T_e^2(t) - T_0^2 \right)
\end{equation}

Changes in the electron-surface scattering rate are also related to the electronic temperature due to a change in the electron occupation in the vicinity of the Fermi level \cite{Stoll2014,Kreibig1995}. If we treat the NPs as spheres or ellipsoids, the change in $\gamma_{e-s}$ is given by

\begin{equation}
\Delta \gamma_{e-s} = \frac{2 v_F}{D_{eq}} \Delta g
\end{equation}

where g is related to the electronic occupation levels, given by

\begin{equation}
g = \frac{1}{E E_F^2} \int_{0}^{\infty} E'^{3/2} \sqrt{E'+E} f(E',T_e)\left[1- f(E'+E,T_e)\right] dE' 
\end{equation}

with $v_F = $1.6 x 10$^6$ m/s for Ag, $D_{eq}$ = $\sqrt{S_{np}/\pi}$, where $S_{np}$ is the surface area of the NP, $E_f$ is the Fermi energy, and f is the Fermi-Dirac distribution.

The change in electron-phonon scattering $\Gamma_{e-ph}$, not the same as the electron-phonon coupling rate, is related to the lattice temperature, and is given by Holstein \cite{Beach1977,Holstein1954} as 

\begin{equation}\label{eq:Holstein}
    \Gamma_{e-ph} = \Gamma_{0}\left[ \frac{2}{5} + 4 \left(\frac{T_L}{\Theta_D} \right)^5 \int_0^{\Theta_D/T_L} \frac{z^4 dz} {e^z-1} \right]
\end{equation}

where $\Gamma_0$ is a parameter determined from Drude fitting $\Theta_D =$ 220 K is the Debye temperature \cite{Owens2010}. The changes in all of these scattering terms result in changes in the Drude term given by

\begin{equation}
\begin{split}
&\Delta\epsilon_{Re}^{Drude} \cong \frac{2 \gamma \omega_p^2}{\omega^4}\Delta\gamma, \\ &\Delta\epsilon_{Im}^{Drude} \cong \frac{\omega_p^2}{\omega^3}\Delta\gamma
\end{split}
\end{equation}
\\
The lattice temperature change modifies the interband dielectric function by shifting the interband edge due to lattice expansion, and is given by \cite{Stoll2014,Antonangeli1974}

\begin{equation}
    \Delta\epsilon_{IB} = -\frac{\partial \epsilon_{IB}}{\partial \omega} \bigg\rvert_\omega \frac{\partial \omega_{IB}}{\partial T_L}\bigg\rvert_{T_0} \Delta T_L
\end{equation}

where the first term is a numerical derivative of the fit the imaginary part of the dielectric constant of Ag \cite{Johnson1972}, and the second term was determined from thermomodulation experiments in bulk \cite{Antonangeli1974} to be between -5.8x10$^{-4}$ and -6.7x1$0^{-4}$ eV K$^{-1}$ for the p to s band transitions.

Finally, the lattice temperature also contributes to changes in the Drude dielectric function by changes in the volume of the particle, which subsequently change the plasma frequency. These changes are given by

\begin{equation}
\begin{split}
&\Delta \epsilon_{Re}^{Drude} \cong \frac{\omega_p^2}{\omega^2}3 \alpha_L \Delta T_L
\end{split}
\end{equation}

where $\alpha_L$ = 1.89 x 10$^{-5}$ K$^{-1}$ is the linear dilation coefficient of Ag. The imaginary part of the Drude change related to the lattice temperature change is given by the change in $\Gamma_{e-ph}$ from Equation \ref{eq:Holstein}.

Now that the changes in both the Drude and interband dielectric functions can be calculated given the changes in the electronic and lattice temperatures from the two-temperature model, these changes are then plugged into a given model for the polarizability of the Ag NP, which depends on the size, geometry, and environmental dielectric functions. To be general, we utilize the expression for the polarizability of an ellipsoidal particle, but generally set two of the in-plane axes (substrate plane) equal, in order to simulate a ``nanoplatelet'' type of structure. After determining the time-dependent differential transmission spectrum for a single Ag NP, the results are then averaged over an inhomogeneous distribution of particle geometries.

\begin{figure}
\centering
\includegraphics[width=0.6\textwidth]{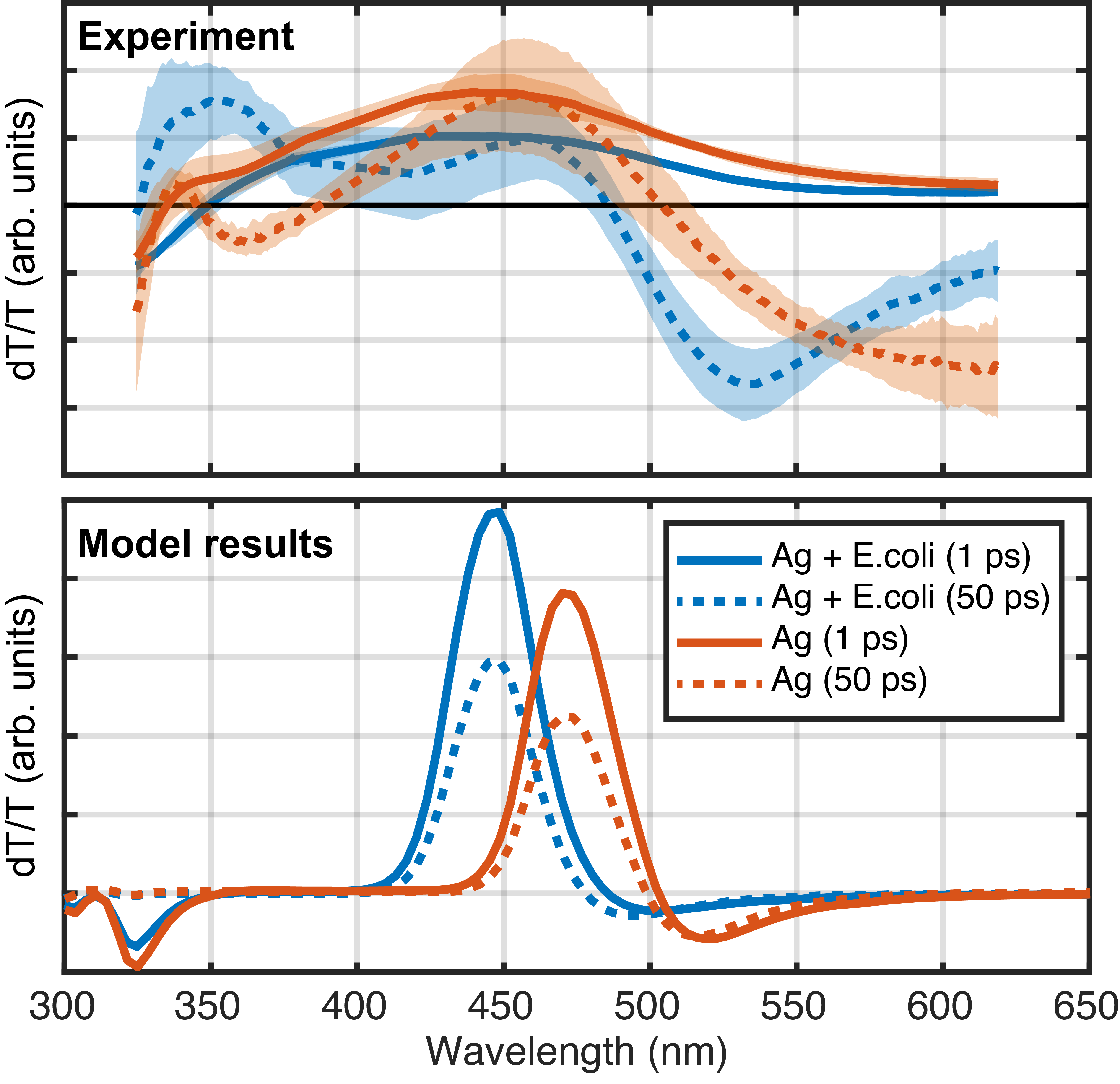}
\caption{Comparison of differential transmission experimental results and modeling calculations. Top: normalized and averaged experimental results  for Ag (orange) and Ag + E.coli (blue) at 1 (solid curve) and 50 (dashed curve) ps. Bottom: modeling results for Ag and Ag + E.coli. Modeling details are described in the main text.}
\label{fig:model}
\end{figure}

The results are displayed in Figure \ref{fig:model}, where the inhomogeneous distribution was chosen to be a Gaussian distribution over the NP thickness with mean of 8 nm and standard deviation of 0.5 nm, consistent with SEM measurements. For Ag pristine at 1 and 50 ps PP delay time, qualitative results are consistent with the experiments: interband absorption modulation is observed around 330 nm, which decays rapidly within the electron-phonon coupling time, indicating that the interband absorption is most sensitive to changes in $T_e(t)$. In the vicinity of the LSPR, the zero-crossing aligns near perfectly with experiment (around 500 nm), and this zero-crossing blue-shifts at longer PP delay times. While the general dispersive lineshape that is indicative of a pump-induced red-shift observed in experiments is reproduced by our modeling, the large linewidth in pristine Ag is not reproduced. We believe that the large signal at long wavelengths requires the full Mie theory treatment, which would incorporate the general nanoplatelet geometry with non-equal axes, as well as quadrupolar resonances that arise due to the large NP size. We also note that this model is unable to reproduce the broad positive features sub 1-3 ps, which may be due to stimulated emission, as discussed earlier in the article. 

Ag + E.coli was also modeled first by changing the environmental dielectric environment to a surface area weighted average of water and glass (also used for Ag pristine, but averaging air and glass), treating the NP like a cylinder, given by

\begin{equation}
\begin{split}
\tilde{\epsilon}_m &= \frac{A_{bottom}\epsilon_{glass}+A_{remain}\epsilon_{top}}{A_{total}}  \\
&= \frac{\epsilon_{glass}+(1+2h/r)\epsilon_{top}}{2+2h/r}
\end{split}
\end{equation}

where h is the height (thickness) of the NP, and r is the radius of the NP. An increase in the environmental dielectric environment leads to a red-shift of the LSPR: thus, some blue-shift mechanism is required to explain the shift in the zero-crossing of Ag + E.coli compared to Ag pristine. Using Equation \ref{eq:deltaEps}, if only the change in the surrounding dielectric and plasma frequency are considered, it is estimated that the plasma frequency increases by around 12\%; this corresponds to a 25\% increase in the electron density. We note that this calculated increase in electron density may seem quite large, requiring either large donation of electrons from the E.coli or significant reduction in the volume of the NP while retaining a constant (or increasing) electron number. Thus, more reasonably, the blue-shift in the LSPR measured in both linear absorption and PP is likely a combination of increase in plasma frequency, ``rounding-out'' of NP geometry, and reduction in NP volume. However, it is shown in Figure \ref{fig:model} that incorporating an increase of 1.7 eV of the plasma frequency ($\hbar \omega_p^0 = $9.1 eV) shifts the zero-crossing of PP, and hence LSPR, to match the experiment. Interestingly, this result is compared to the calculations of Henglein \cite{Henglein1993} who calculated the effect of adsorption of nucleophiles, and hence donation of electrons, to Ag NPs. There, they showed that assuming electrons were only donated to the surface of the NP that the Fermi energy would increase by around 8\%, qualitatively consistent with the percentage increase of 1.7/9.1 = 18\% observed here. Henglein also noted that this treatment may be naive, in the sense that the ``donated'' electrons are still bound to the nucleophile, and may only contribute to the modification of the Ag dielectric by a reduction in the effective NP volume, leading to an increase in electron density.

We believe that the modeling approach presented here may be improved in a number of ways. First, we note the limitations of the model. The ratio of the amplitude of the pump-induced absorption at 500-600 nm to the positive bleaching feature at 460 nm produced in our calculations does not match the experiment. Second, the widths of the dispersive features in both Ag and Ag+E.coli do not match the experiment. We believe that both of these problems might be solved by utilizing the full Mie scattering theory, which would take into account the quadrupole mode, as well as features that arise in high aspect ratio particles at longer wavelengths.

\bibliography{library}